\documentclass[preprint]{aastex}
\usepackage{natbib}
\usepackage{amsmath, amsthm, amssymb}
\usepackage{subfigure}
\usepackage{graphicx}

\usepackage{natbib}
\usepackage{hyperref}
\usepackage{lscape}
\usepackage{pdflscape}

\newcommand{\ltsima} {$\; \buildrel < \over \sim \;$}
\newcommand{\gtsima} {$\; \buildrel > \over \sim \;$}
\newcommand{\lta} {\lower.5ex\hbox{\ltsima}}
\newcommand{\gta} {\lower.5ex\hbox{\gtsima}}

\newcommand{\ffm}[1]{}
\newcommand{\ffc}[1]{}
\newcommand{\fpm}{}
\newcommand{\fpc}[1]{}

\def\mod{}

\def\apj{Ap. J.}

\def\apjl{Ap. J. Lett.}
\def\apjs{Ap. J. Supp.}

\def\mnras{Mon. Not. RAS}
\def\physrep{Physics Reports}
\def\aap{Astron. \& Astrophys.}

\def\nat{Nature}

\slugcomment{Date: \today; Accepted for publication in ApJ}

\begin{document}

\title{{\it Fermi} detection of delayed GeV emission from the short GRB~081024B}
\author{
A.~A.~Abdo\altaffilmark{2,3}, 
M.~Ackermann\altaffilmark{4}, 
M.~Ajello\altaffilmark{4}, 
K.~Asano\altaffilmark{5,1}, 
W.~B.~Atwood\altaffilmark{6}, 
M.~Axelsson\altaffilmark{7,8}, 
L.~Baldini\altaffilmark{9}, 
J.~Ballet\altaffilmark{10}, 
G.~Barbiellini\altaffilmark{11,12}, 
D.~Bastieri\altaffilmark{13,14}
% ...
}
\altaffiltext{1}{Corresponding authors: K.~Asano, asano@phys.titech.ac.jp; P.~N.~Bhat, narayana.bhat@nasa.gov; N.~Omodei, nicola.omodei@gmail.com.}
\altaffiltext{2}{Space Science Division, Naval Research Laboratory, Washington, DC 20375, USA}
\altaffiltext{3}{National Research Council Research Associate, National Academy of Sciences, Washington, DC 20001, USA}
\altaffiltext{4}{W. W. Hansen Experimental Physics Laboratory, Kavli Institute for Particle Astrophysics and Cosmology, Department of Physics and SLAC National Accelerator Laboratory, Stanford University, Stanford, CA 94305, USA}
\altaffiltext{5}{Interactive Research Center of Science, Tokyo Institute of Technology, Meguro City, Tokyo 152-8551, Japan}
\altaffiltext{6}{Santa Cruz Institute for Particle Physics, Department of Physics and Department of Astronomy and Astrophysics, University of California at Santa Cruz, Santa Cruz, CA 95064, USA}
\altaffiltext{7}{Department of Astronomy, Stockholm University, SE-106 91 Stockholm, Sweden}
\altaffiltext{8}{The Oskar Klein Centre for Cosmoparticle Physics, AlbaNova, SE-106 91 Stockholm, Sweden}
\altaffiltext{9}{Istituto Nazionale di Fisica Nucleare, Sezione di Pisa, I-56127 Pisa, Italy}
\altaffiltext{10}{Laboratoire AIM, CEA-IRFU/CNRS/Universit\'e Paris Diderot, Service d'Astrophysique, CEA Saclay, 91191 Gif sur Yvette, France}
\altaffiltext{11}{Istituto Nazionale di Fisica Nucleare, Sezione di Trieste, I-34127 Trieste, Italy}
\altaffiltext{12}{Dipartimento di Fisica, Universit\`a di Trieste, I-34127 Trieste, Italy}
\altaffiltext{13}{Istituto Nazionale di Fisica Nucleare, Sezione di Padova, I-35131 Padova, Italy}
\altaffiltext{14}{Dipartimento di Fisica ``G. Galilei", Universit\`a di Padova, I-35131 Padova, Italy}
% ...

\begin{abstract}
\mod{We report on the detailed analysis of the high-energy extended emission from the short Gamma-Ray Burst (GRB) 081024B, detected by the {\it Fermi} Gamma-ray Space Telescope. Historically, this represents the first clear detection of temporal extended emission from a short GRB.}
The light curve observed by the {\it Fermi} Gamma-ray Burst Monitor lasts approximately 0.8 seconds whereas the emission in the {\it Fermi} Large Area Telescope lasts for about 3 seconds. 
Evidence of longer lasting high-energy emission associated with long bursts has been already reported by previous experiments. Our observations, together with the earlier reported study of the bright short GRB 090510, indicate similarities in the high-energy emission of short and long GRBs and open the path to new interpretations. 
\end{abstract}


\section{Introduction}

Gamma-Ray Bursts (GRBs) are extremely energetic and brief explosions originating at cosmological distances. Since their discovery in the soft gamma-ray regime roughly 40 years ago, they have been detected to emit in almost every wavelength, from mm to the GeV range. The properties of their gamma-ray emission (e.g., duration, spectral shape, variability) have been thoroughly studied in the past 20 years, with multiple spacecrafts, notably with
the Burst And Transient Source Experiment ({\it BATSE}) onboard the Compton Gamma-Ray Observatory ({\it CGRO})~\citep[see][for a review]{zhangmes04}. 
One of their most enduring properties is their classification in two duration classes \citep{kouvel93}, with distinct spectral characteristics: short (\ltsima 2 s) hard GRBs and long  soft ones. 
This bimodality has been confirmed with NASA's {\it Swift} satellite~\citep{swift} and recently with the Gamma-ray Burst Monitor \citep[{\it GBM},][]{meegan09} onboard the {\it Fermi} satellite. The prevailing notion is that these two GRB classes originate from different progenitor systems.

\mod{The first detection of an X-ray afterglow goes back to observation of GRB 970217 with the {\it BeppoSax} satellite \citep{costa97}, followed by the discovery of the optical transient that allowed the first ever determination of the redshift for a GRB \citep{vanPara97}}.
Multi-wavelength (radio to X-ray) follow-up observations of GRB afterglows and the extensive study of the GRB hosts and their environments, have now established
that at least some long GRBs are connected with the collapse of massive rapidly rotating stars into black holes \citep[][and references therein]{wb06}. The
origin of short GRBs is less certain. In the last few years growing evidence supports the idea that short GRBs originate from neutron star-neutron star (NS-NS)
mergers also ending in a few solar mass black-hole surrounded by a short-lived accretion disc \citep[see][]{nakar07,eichler89,
  rufjan99,rosswog00,narayan01,rosswog03}. It is, however, rather difficult observationally to distinguish mergers from collapsars. SN explosions have been spectroscopically associated in a scant four cases with GRBs, but none of these has been a short GRB.

The detection of high-energy ($\gtrsim$100~MeV) prompt and afterglow emission from GRBs with {\it Fermi} offers a unique probe to test the properties of the outflow, determine the mechanism of energy transfer, and to address fundamental physics issues like Lorentz invariance, as has been done for other bursts such as GRB~080916C \citep{GRB080916C,zhangb09} and GRB~090510~\citep{GRB090510}. During its first year of operation {\it Fermi} detected high-energy photons from two short GRBs \citep{gcn081024b,gcn090510}, opening a new window in our understanding of the GRB phenomenon. 

High-energy emission above 1~GeV had been detected in the past with {\it CGRO}/EGRET \citep{sommer94} in association with several long {\it BATSE}
GRBs. Little was known, however, about the radiation physics of short GRBs, due to poor statistics at higher energies \citep{kaneko06}. 
A notable exception is the extremely bright GRB~930131~\citep{kouvel94}, which consisted of an initial very intense complex of pulses with total duration about 1.5 second, superimposed on a significant shallow tail lasting up to $\sim$50 s. As the tail fluence was at least an order of magnitude lower than the one of the initial complex, it was the initial short pulse that dominated the event energetics and most likely contained most of the burst flux. Apart from its intensity, GRB~930131 was also remarkable in its high-energy properties; {\it EGRET} detected 16 high-energy gamma rays, 
including the highest energy photon of 1.2 GeV 26.5 s after the initial spike. As the event was not classified as a short GRB, however, it was not clear until
recently that short bursts could produce such high-energy emission, and whether the properties of the short GRB emission resembled those of the long events at GeV energies.

\mod{In this paper, we report on the observation and the analysis of the GeV emission from the short GRB~081024B.}
The event lasts approximately 0.8~s below 5~MeV, entirely consistent with the short GRB class \citep{kouvel93}. The emission above 100~MeV lasts for about 3 seconds, thus showing 
unambiguous evidence for a delayed high-energy component in a short GRB. 

\section{Observations}

The {\it Fermi} Gamma-ray Space Telescope was launched on 2008 June 11, and during the first year of science operations detected high-energy emission from 9 GRBs\footnote{\url{http://fermi.gsfc.nasa.gov/ssc/resources/observations/grbs/grb_table/}}, doubling the number of bursts detected above 100 MeV. 
Two instruments operate onboard {\it Fermi}: the Gamma-ray Burst Monitor (GBM)~\citep{meegan09} covering the energy range from 8~keV to 40 MeV, and the Large Area Telescope (LAT)~\citep{atwood09}, from 20 MeV to more than 300~GeV. 

On 2008 October 24 at 21:22:41 (UT) the LAT detected an increased count rate associated with the short burst GRB~081024B, which
  triggered the GBM~(trigger number 246576161, \citep{gcn081024b-GBM}). 
The LAT ground analysis for burst detection and localization follows the procedure described in~\cite{GRB080825C}. Selecting ``transient'' events with energy above 100 MeV, 
final localization is found to be RA = 322{$^\circ$}.9 
  , Dec = 21{$^\circ$}.2 
  , with a statistical uncertainty of 0{$^\circ$}.2 (68\% containment radius). 
\mod{The map of the Test Statistics reaches its maximum value, TS$_\mathrm{max}$ = 48.9, at this position, corresponding to a 6.7~$\sigma$ detection.
We also estimated the significance of the high-energy emission using the Li \& Ma method \citep{LiMa83}, which computes the probability of the temporal excess in the case of Poisson statistics. We used two different event selections, considering a fixed region of interest (ROI) of 15$^\circ$, and an energy dependent ROI, taking into account the energy dependence of the LAT point spread function (PSF).
The LAT recorded 11 events with reconstructed energy above 100~MeV, within 15$^\circ$ from the position of the burst and within 3~s from the trigger time, when the expected number of counts from the background is 0.75, corresponding to an excess of 6.8~$\sigma$. The number of expected counts has been computed considering the background 100 s before and 100 s after the burst, excluding the time window of 3 s during the burst. A more careful estimation of the background, using Monte-Carlo simulations to estimate the charged particle background, and six months of data for estimating the gamma-ray background was also performed, providing a significance of 6.7~$\sigma$. Details of the method are described in \cite{GRB080825C}. 
If we consider the energy dependent ROI, considering only the events that are within three times the 68\% containment radius of the PSF (approximately corresponding to the 95\% of the containment radius), the number of expected events from background decreases to 0.08,  and the significance of the excess increases up to 8.5~$\sigma$. This higher significance can be easily understood as the LAT \fpm{PSF} strongly depends on the energy ($\propto E^{-0.8}$), which makes an energy dependent ROI very effective in reducing the background contamination, without loosing gamma-ray events from the source.}

{\mod The LAT position} was observed by the Swift X-Ray Telescope (XRT), starting 70.3 ks after the GRB trigger time and lasting 9.9 ks, but no X-ray counterpart was found \citep{gcn081024b-XRT}.
An additional 13.5 ks observation was conducted the next day \citep{gcn081024b-XRT-2} and confirmed the lack of an X-ray afterglow candidate.
Optical observations also produced no counterpart \citep{gcn081024b-P200}.
The absence of an XRT detection so long after a short burst is not too surprising.  A study of the Swift XRT catalog
of GRBs\footnote{\url{http://swift.gsfc.nasa.gov/docs/swift/archive/grb_table}}
 shows that of the 33 short GRBs for which XRT observations were made, only 10 had detectable flux 75 ks after the trigger time.

\section{Light curves, durations and spectral lags}

\begin{figure}
  \epsscale{1}
  \plotone{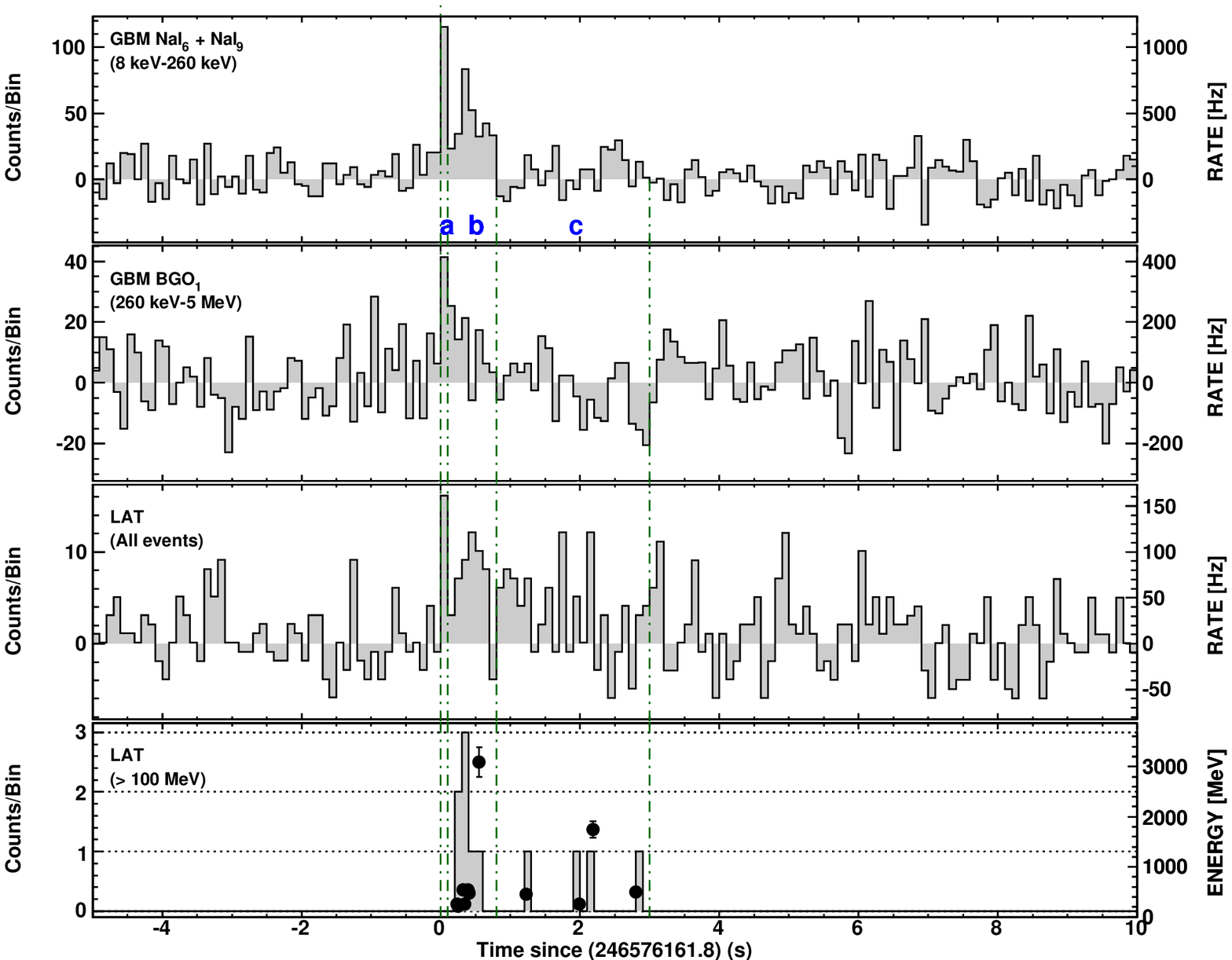}
  \caption{Multi-instrument light curve for GRB~081024B. The top panel shows the sum of the \fpm{background subtracted} signal from two NaI detectors. The second panel is one BGO \fpm{detector}. The third panel shows all the events recorded by the LAT, without any selection on the quality of the events (background subtracted). The fourth panel shows the selected ``transient'' events above 100 MeV. The energy of events is reported at the right axis of the plot. The arrival times, the reconstructed positions and the energies of the 11 events are reported in Table~\ref{table:events}.}
  \epsscale{1}
  \label{fig:lightcurve}
\end{figure}

The multi detector light curve is shown in Fig.~\ref{fig:lightcurve}. 
The top panel shows the background subtracted light curve for the summed signal of the two brightest NaI detectors (6 and 9) between 8~keV and 260~keV. 
The background subtracted light curve of the brightest BGO detector (1) is shown in the second panel (260 keV -- 5 MeV). 
The third panel shows the LAT signal without any selection (i.e., all the events that passed the onboard gamma filter). 
The quality of these events is not good enough to use them in the spectral analysis, but the properties of the ensemble can be assessed quantitatively by means of a dedicated Monte-Carlo simulation and do convey physical information in the extremely interesting energy range in which the BGO and the LAT overlap.
These three light curves are characterized by a narrow spike of about 0.1~s (interval ``a''), followed by a longer pulse, of about 0.7~s (interval ``b''). There is no evidence of emission after $\sim$0.8~s in the NaI and BGO detectors.
The \mod{bottom} panel shows the light curve of the ``transient" selected events with well defined direction and energy (above 100 MeV). The arrival times and the reconstructed energies (right axis) of the selected events are also displayed. 
We estimate a total of 0.4 background events during the time interval shown in Fig.~\ref{fig:lightcurve}.
An event with energy $3.1\pm0.2$~GeV was detected after 0.55~s while a second event 
of $1.7\pm0.1$ GeV was detected after 2.18~s.
Table~\ref{table:events} contains the arrival times, the energies with the estimated error, and the arrival directions of these eleven selected events. The last two columns of the table are the estimated 68\% containment radii calculated from the point-spread function (PSF) and the distance from the localization of the GRB, in PSF units.

\mod{We studied the narrow spike visible in the full lightcurve (third panel) of interval ``a''. The probability to obtain the same number of counts from background fluctuations is discarded at the 3.5~$\sigma$ level. Furthermore,} we performed dedicated Monte-Carlo simulations to estimate properly the energy of these LAT photons.
These events do not belong to the ``transient'' class, which is the most generous event selection with minimal requirements on direction and energy reconstructions. Typically, they are discarded because they produce very few hits in the tracker ($<$20), with a very short track, and very low raw energy deposited in the calorimeter ($<$5 MeV). 
These topologies are typical of low-energy events, with energies between 10 MeV and 40 MeV. 
\mod{If we select these topologies in the data, the probability that this narrow pulse is the result of a background fluctuation decreases to 5.9~$\sigma$ level.}

We conclude that the spike in interval ``a'' in the LAT data is \mod{significant,} with energies below 100~MeV (although the energy resolution for this class of events is relatively poor. $\sim$50\%, from Monte-Carlo simulations). 

A common method for estimating GRB durations is to compute the T$_{90}$ which measures the duration of the time interval during which 90\% of the total observed counts have been detected~\citep{kouvel93}. 
Background fluctuations, especially in weak GRBs, strongly affect T$_{90}$'s as often the fluctuations are comparable to the 5\% of the total GRB fluence. Values of T$_{50}$, however, are generally more robust, and especially so in the case of weak events. 
For GRB~081024B, the T$_{50}$ (T$_{90}$) is 0.33~s (0.66~s) in the NaI detectors, 0.15~s (0.27~s) in the BGO, while it is significantly longer for the LAT, corresponding to 0.9~s (2.1~s) for the full statistic light curve and 1.5~s (2.6~s) selecting only the events above 100 MeV.  

Spectral lags are characteristic of long GRBs, which exhibit hard-to-soft 
spectral evolution, while short GRBs do not exhibit such a property~\citep{norris06}. 
We searched for a possible spectral lag in GRB~081024B using the cross-correlation function (CCF).
The GBM Time Tagged Event light curves of the 4 brightest NaI detectors (6, 7, 9 and 10) were summed with 100~ms time resolution in 8 logarithmic energy bins from 8 to 1950~keV. 
Similarly, the two BGO light curves were added in 8 bins from 0.11 to 107.6 MeV. 
For the LAT, we used all photons above 100~MeV.  
The errors on the CCFs were estimated using Bartlett's formula~\citep{bartlett78} and were propagated to the errors of the peak position.
We computed the CCFs between the GBM/NaI and the GBM/BGO detectors as well as
between the GBM and the LAT. 
We found no energy-dependent delay between any of the data types used. We also used two resolutions (50 and 100 ms) and verified that our results did not change significantly.

\section{Spectral Analysis}

We performed a time-resolved spectral analysis in intervals ``a'', ``b'', and ``c'' of Fig.~\ref{fig:lightcurve}, simultaneously fitting the signal from the two
NaI (selecting all the channels between 8~keV and 860~keV), the BGO (from 200~keV to 36~MeV) and the LAT detectors (selecting ``transient'' events above
100~MeV)\footnote{Here we have used the post-launch Pass 6 v3 IRFs, which take into account, on average, the correction of the inefficiency due to pile-up of
  cosmic rays. The version of Science Tools used is v15r9p2. We have used the package {\em rmfit} (version 3.1) to perform the GBM/LAT joint fit.}.
Table~\ref{table:fitresults} shows the results for all time intervals, testing different fitting functions. 



In interval ``a'' the best fit to the GBM data is obtained with a power-law with exponential cutoff~\citep[see COMPT model]{kaneko06}.
The best fit parameters are summarized in Table~\ref{table:fitresults} as fit~``1'' .  
The peak energy $E_\mathrm{peak}$ lies in the BGO energy range, and, even though its value is only marginally constrained, is consistent with a very hard
spectrum, with a roll-off at energies above a few MeV. 
The LAT upper limit on the photon flux in the 100 MeV--10 GeV energy range is 
$3.8\times 10^{-5}$ ph cm$^{-2}$ s$^{-1}$ or, in energy flux, $4.7\times 10^{-9}$ erg cm$^{-2}$ s$^{-1}$, and is consistent with the extrapolated flux from the COMPT function fitted to the GBM data.
We also performed a fit using the Band function~\citep{band93}. The resulting parameters are listed in Table~\ref{table:fitresults} as fit~``2''. 

Interval ``b'' is best represented by a Band function with the parameters displayed in Table~\ref{table:fitresults} as fit~``3''. 
To estimate the significance of the spectral evolution from interval ``a'' to interval ``b'' we fit the count spectrum in interval ``a'' assuming a Band function with a fixed high-energy spectral index obtained by the best fit of interval ``b'' ($\beta$ = -2.1), and we estimate the number of expected events in the LAT detector. 
Based on several realizations, the average number of expected events in interval ``a'' is 2.6, for a chance probability of observing zero counts of about 7\%. 
Therefore, we can only suggest that a spectral evolution characterizes the temporal behavior of this burst, somewhat similar to that observed in the first portion of the emission
of the long, bright GRB 080916C~\citep{GRB080916C}. 
Fit~``4'' shows the result of the fit when a COMPT model plus a power-law is adopted.
Even though the fit is not favored, and the statistics are very low for any conclusive remark, this model has some interesting implications, such as the possible presence of an extra component in LAT data.

\begin{figure}
  \epsscale{1}
    \plotone{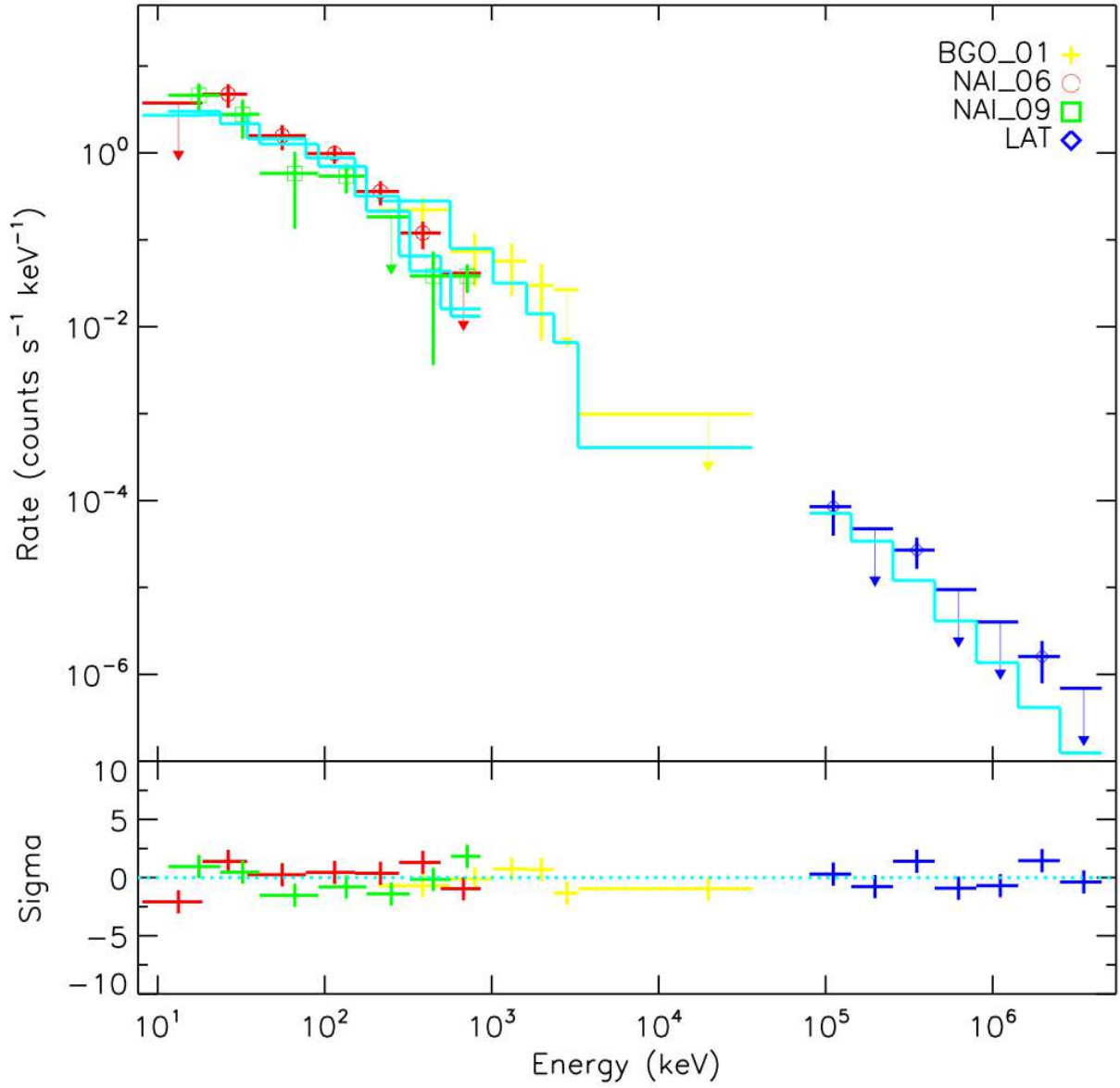}
  \caption{Count spectrum in interval ``b'', fit ``3'' . The data points are the rates from GBM (red circles: NaI$_6$, green squares: NaI$_9$, yellow crosses: BGO$_1$) and LAT (blue diamonds). The predicted rates  in the various detectors are obtained by folding the best fit model \fpm{(Band function)} with the response of the detectors, and are displayed as continuous cyan lines.}
\epsscale{1}
\label{fig:Spectrum1}
\end{figure}

\begin{figure}
  \epsscale{0.8}
      \plotone{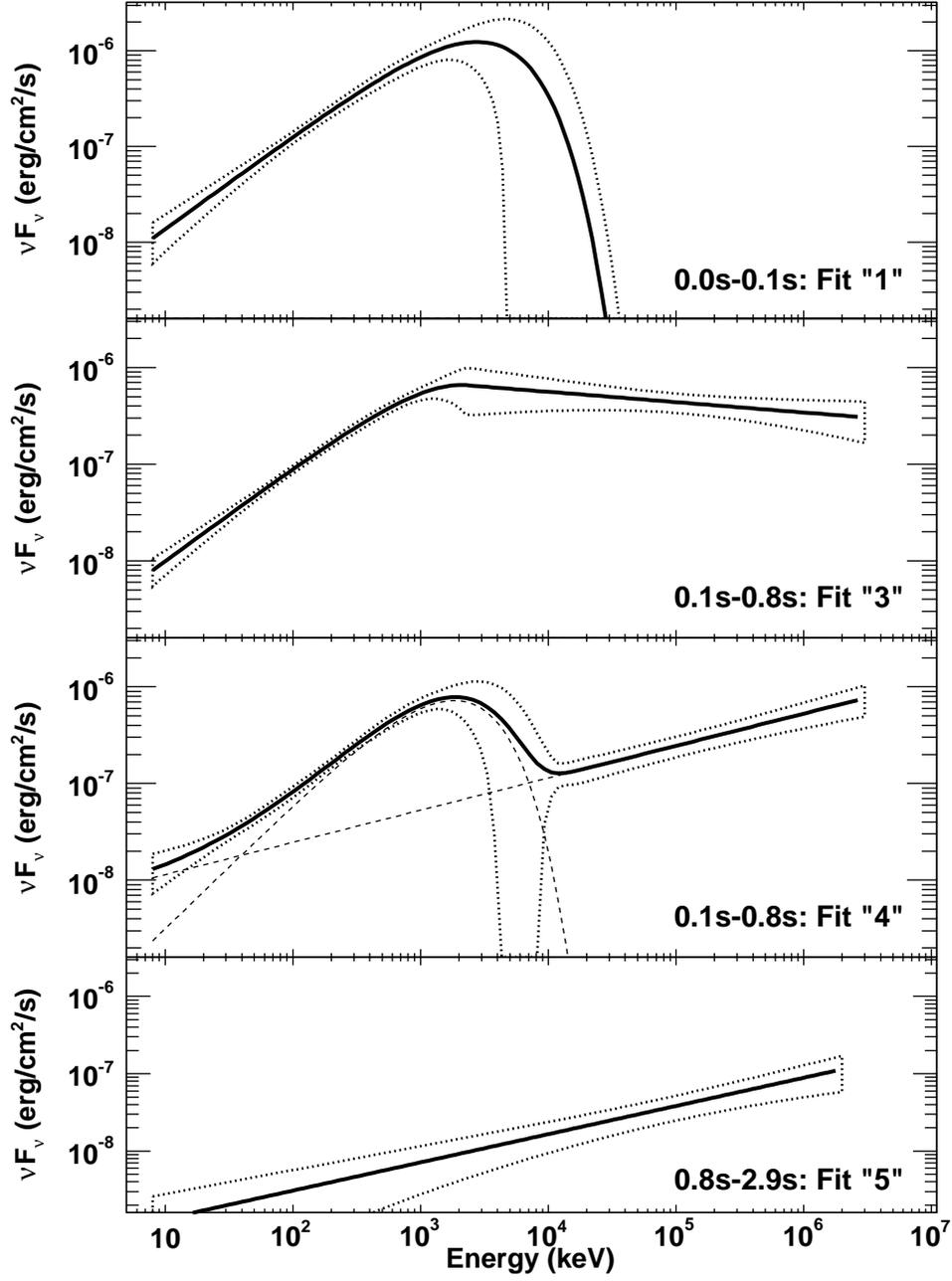}
 \caption{Analytical representation of the source photon spectra. From top to bottom: fit ``1'' in interval ``a'' (COMPT model); fit ``3'' in interval ``b'' (Band); fit ``4'' in interval ``b'' (COMPT plus power-law); fit~``5'' in interval ``c'' (power-law).}

\epsscale{1}
\label{fig:Spectrum2}
\end{figure}

The last entry in the table is related to interval ``c'' which is best represented by a simple power-law. Fig.~\ref{fig:Spectrum1} shows the count spectrum for the Fit``3'' 
and  Fig.~\ref{fig:Spectrum2} the stacked $\nu F_{\nu}$ plot 
for Fit 1,3,4 and 5, where the 68\% CL are computed from the covariance matrix provided by the fitting routine ({\it rmfit}).

Systematic errors are due to uncertainties in the effective areas of the different detectors, energy resolution and background estimation. The most
important contribution arises from the uncertainties related to the effective areas. For the LAT they have been derived from a study of the Vela pulsar
\citep{VelaPSR} and are 10\% below 100 MeV, 5\% around 1 GeV and 20\% above 10 GeV. We adopt a 10\% uncertainty in the NaI and BGO effective area (both overall
normalization and slope). We propagate the systematic errors on the effective areas to the parameters of the model, and we compute the systematic errors associated to the flux/fluence values. Systematic errors in each case are comparable or smaller than the statistical errors quoted in Table~\ref{table:fitresults}.

\section{Discussion and interpretation}
GRB~081024B was the first short GRB with observed emission above GeV energies.
Unlike GRB~930131~\citep{kouvel94,sommer94},
a 3 GeV photon from GRB~081024B is well correlated with the second low-energy pulse.
The cross-correlation function of the  light curves between 30-100 keV and 100-300 keV shows no strong signature of spectral lag larger than 30 ms,
which is consistent with the negligible spectral lags in other short GRBs~\citep{norris06}.

While the majority of long GRB spectra are well fitted by the conventional Band function, 
previous spectral analyses of short GRBs have mostly used the cutoff power-law function \citep{ghirlanda04,mazets04}.
The exponential cut-off implies that the bulk motion of short GRBs is not necessarily
ultra-relativistic, owing to the compactness problem for high-energy photons
above $m_{\rm e} c^2$ \citep{meszaros02}, which becomes less severe (see~\citep{nakar07} 
for a quantitative estimate of the $\Gamma_{\rm min}$ in this case).
This difference between long and short GRBs may be due to poor counting statistics at high energies in short GRBs, stressing the need for a larger sample with sufficient high-energy
photons in MeV-GeV bands.

The delayed onset of a GeV pulse, which is frequently found in
other LAT-detected bursts such as GRB 080916C or GRB 080825C, may be explained
by the different physical parameters for two pulses \citep{GRB080916C},
$\gamma \gamma$ pair-production opacity effect \citep{granot08},
or acceleration timescale of high-energy protons for
hadronic models
\citep[e.g.,][]{rachen98, dermer02,razzaque05, dermer06, asano07, asano09}.
The long-lasting tail of GeV emission is also common to GRB 080916C, GRB 080825C and GRB 090510.
One possible interpretation is that the long tail is synchrotron or
synchrotron self-Compton (SSC) emission during the afterglow phase~\citep{ghi09}.
Alternatively, the GeV afterglow emission may originate from cascades
induced by ultrarelativistic hadrons accelerated by the blast wave~\citep{bottcher1998}.
In these afterglow scenarios even the 3 GeV photon at $t \simeq 0.5s$ could have an afterglow origin, and the delayed onset of GeV emission is also naturally explained.
The onset time \citep{molinari07} and the hard spectrum for interval
ÒcÓ do not contradict the afterglow scenarios.
Early afterglow models for this long lasting tail with synchrotron emission \citep{he09} and
SSC emission \citep{corsi09} are actually proposed.
On the other hand, \cite{corsi09} pointed out that
SSC emission from late internal shocks can be an alternative interpretation
for the long lasting tail.

\begin{figure}
  \epsscale{1}
  \plotone{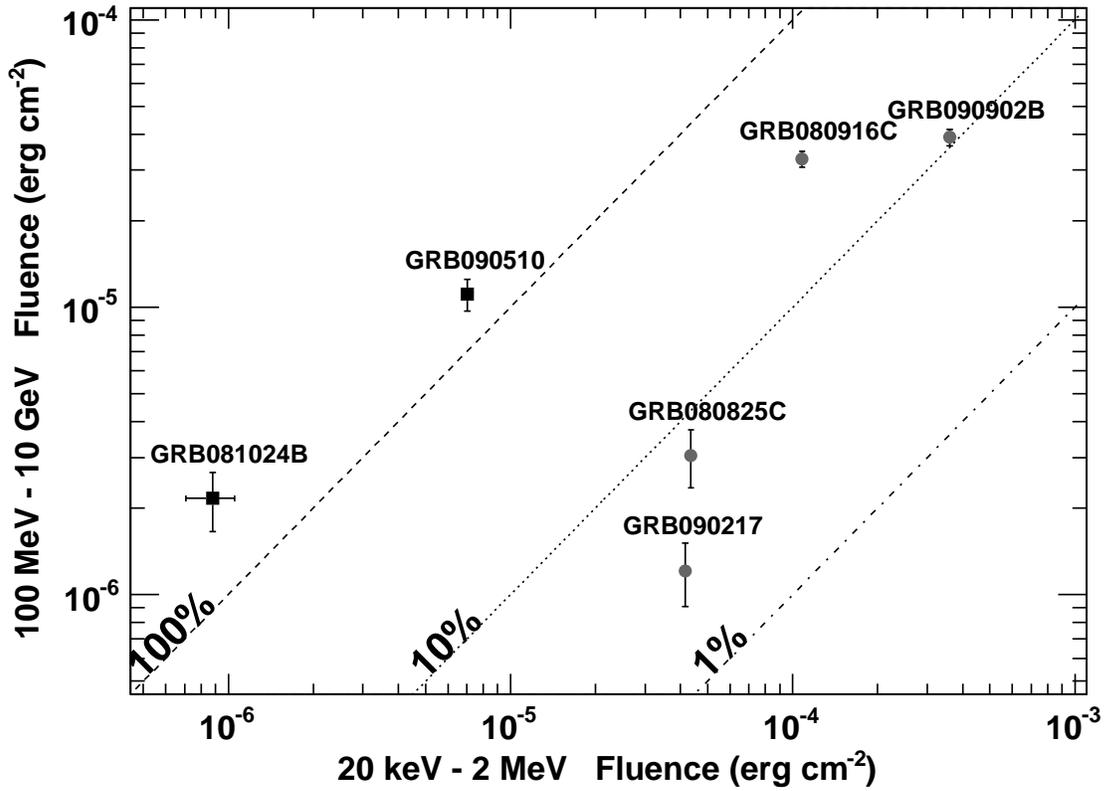}
  \caption{Fluence-fluence diagram for 6 LAT GRBs. Black squares are classified as short GRBs and filled grey circles represent long GRBs. The dashed lines indicate the 1, 0.1, 0.01 values of the ratios between the 100 MeV-10 GeV fluence and the 20 keV-2 MeV fluence.}
  
  \epsscale{1}
  \label{fig:fluflu}
\end{figure}

Finally, in Fig.~\ref{fig:fluflu} we show the high-energy fluence versus low-energy fluence diagram, as suggested by~\citep{le09}, comparing 6 LAT GRBs. This is considered as a way to search for separate classes of GRBs and, specifically, spectral differences between the short-hard and long duration GRB classes. The sample is limited and therefore no definite conclusion can be drawn, but we note that the two short bursts so far detected by the LAT 
occupy a region where the energy emitted at high energy is greater than the energy emitted at lower energy (with a ratio $\gta$1) suggesting that short GRBs may have an higher efficiency in emitting $\gamma$-rays. Nowadays only two short GRBs have been detected with a significant emission at GeV energies; we expect that in the next decade we will be able to significantly increase the {\it Fermi} short GRB sample and understand their similarities and differences from the long, soft GRBs.

\acknowledgments
The $Fermi$ LAT Collaboration acknowledges support from a number of agencies and institutes for both development and the operation of the LAT as well as scientific data analysis. These include NASA and DOE in the United States, CEA/Irfu and IN2P3/CNRS in France, ASI and INFN in Italy, MEXT, KEK, and JAXA in Japan, and the K.~A.~Wallenberg Foundation, the Swedish Research Council and the National Space Board in Sweden. Additional support from INAF in Italy and CNES in France for science analysis during the operations phase is also gratefully acknowledged.

\hyphenation{Post-Script Sprin-ger}

\begin{deluxetable}{c c c c c c c} 
\tabletypesize{\scriptsize}
\tablecaption{Selected LAT  ``transient'' events with energy above 100 MeV.}
\tablewidth{0pt}
\tablehead{
 & \colhead{Time-T$_\mathrm{trig}^{1}$} & \colhead{Energy$^{2}$} & \colhead{RA}      & \colhead{Dec}      & \colhead{PSF$^{3}$} & \colhead{Dist.$^{4}$} \\
 & \colhead{s}                                           & \colhead{MeV}               & \colhead{degrees} & \colhead{degrees} & \colhead{degrees}       & \colhead{in PSF}\\
}
\startdata
        1 & 0.229 & 145 $\pm$ 22      & 325.3 & 21.4 & 2.6 & 0.8\\
        2 & 0.248 & 101 $\pm$ 17       & 323.2 & 18.7 & 3.4 & 0.7\\
        3 & 0.320 &  442 $\pm$ 44      & 322.6 & 20.6 & 1.1 & 0.6 \\
        4 & 0.342 & 140 $\pm$ 21     & 322.2 & 20.2 & 2.7 & 0.4 \\
        5 & 0.391 & 441 $\pm$ 44     & 323.6 & 20.9 & 1.1 & 0.7 \\
        6 & 0.406 & 368 $\pm$ 41     & 323.1 & 18.8 & 1.2 & 1.2 \\
       {\bf 7} & {\bf 0.551 } & {\bf 3070 $\pm$ 230}  & {\bf 322.9} & {\bf 21.2} & {\bf 0.2} &{\bf 0.2}\\
        8 &  1.223 & 350 $\pm$ 39    & 324.1 & 20.7 & 1.3 & 1.0 \\
        9 & 1.986 & 143 $\pm$ 22  & 325.0 & 24.2 & 2.6 & 1.4 \\
       {\bf 10} & {\bf 2.184} & {\bf 1680 $\pm$ 130}  & {\bf 322.7} & {\bf 21.5} & {\bf 0.4}& {\bf 1.0} \\
       11 & 2.801 & 386 $\pm$ 41 & 322.6 & 20.1 & 1.2 & 1.0 \\
\enddata
\tablecomments{$^{1}$ Arrival time with respect to the GBM trigger time T$_\mathrm{trig}$ = 246576161.864; $^{2}$ reconstructed energy with estimated error. \mod{Errors are estimated using Montacarlo simulations from the width of the reconstructed energy distribution}; $^{3}$ evaluated 68\% containment radius from the PSF for ``transient'' events.$^{4}$ distance from the GRB position, in PSF units.}
\label{table:events}
\end{deluxetable}

\clearpage
\begin{landscape}

\begin{deluxetable}{clccccccccc}
\setlength{\tabcolsep}{0.04in}
\tabletypesize{\scriptsize}
\tablecaption{Fit parameters for the  time-resolved spectral fits.}

\tablewidth{530pt}
\tablehead{
\colhead{Fit} & \colhead{Interval} & \colhead{Detectors} & \colhead{Model} & \colhead{E$_\mathrm{peak}$} & \colhead{$\alpha$} & \colhead{$\beta$} & \colhead{C-Stat/DOF} & \colhead{Fluence} & \colhead{Fluence} \\ 
 &  &  &  &   &  &  &  & (20 keV -- 2 MeV)  & (100 MeV -- 10 GeV)\\ 
& \colhead{(s)} &   & & \colhead{(MeV)}  & & & &  \colhead{erg/cm$^2$} & \colhead{erg/cm$^2$} \\
}
\startdata
1 & ``a'', 0--0.1  & NaI+BGO &  COMPT  & $2.7^{\,+4.3}_{\,-1.5}$ &  $-1.03^{\,+0.23}_{\,-0.19}$ & -  & 330/352 & $(1.7\pm0.3)\times 10^{-7}$ & $<4.7\times 10^{-10}$ \\
2 & ``a'', 0--0.1 & NaI+BGO &  BAND     & $2.8^{\,+5.0}_{\,-1.5}$ &  $-1.03^{\,+0.23}_{\,-0.20}$ & $<-1.7^{\dag}$ & 330/351 & $(1.7\pm0.4)\times 10^{-7}$  & $<7.3\times 10^{-8}$ \\
\hline
3 & ``b'', 0.1--0.8 &NaI+BGO+LAT &  BAND  & $2.0^{\,+1.9}_{\,-1.0}$ &  $-1.03^{\,+0.17}_{\,-0.14}$ & $-2.10^{\,+0.11}_{\,-0.14}$ & 383/359& $(7.9\pm1.9)\times 10^{-7}$ &  $(8.9\pm3.0)\times 10^{-7}$  \\
4 & ``b'', 0.1--0.8 & NaI+BGO+LAT &  COMPT + & $1.6^{\,+1.5}_{\,-0.6}$ &  $-0.7^{\,+0.4}_{\,-0.3}$ & - & &  &  \\
&        &        &         POW      &  - &  -  & $-1.68^{\,+0.10}_{\,-0.06}$ &  384/358 & $(8.8\pm2.0)\times 10^{-7}$ &  $(1.1\pm0.4)\times 10^{-6}$ \\
\hline
5 & ``c'', 0.8--2.9 & NaI+BGO+LAT & POW & - &  - & $-1.6^{\,+0.4}_{\,-0.1}$ & 274/361 & $(4.3\pm3.2)\times 10^{-8}$ & $(4.0\pm2.4)\times 10^{-7}$ \\
\hline
\enddata
\tablecomments{For each time bin (``a'', ``b'' and ``c''), and for each fit (1 to 5), we report the detectors used in the fit, the name of the model used, and \fpm{the value of each spectral parameter}. Asymmetric errors are obtained from the profile of the Cash statistic. The pivot energy of the power-law function is set to 100~MeV.$^{\dag}$The upper limit for $\beta$ has been obtained looking at the profile of the Cash statistic. Although in this fit $\beta$ is basically undetermined, we report it for reference.}
\label{table:fitresults}
\end{deluxetable}

\end{landscape}
\clearpage

\end{document}